# Design and test results of scientific X-ray CMOS cameras

Wenxin Wang[a], Zhixing Ling*[a], Chen Zhang[a], Qiong Wu[b], Zhenqing Jia[a], Xinyang Wang[c], Weimin Yuan[a], Shuang-Nan Zhang[a,d]

[a]National Astronomical Observatories, Chinese Academy of Sciences, Beijing, 100101, China; [b]Department of Engineering Physics, Tsinghua University, Beijing, 100084, China; [c]Gpixel inc., Changchun, 130033, China; [d]Institute of High Energy Physics, Chinese Academy of Sciences, Beijing,100049, China

## ABSTRACT

In recent years, scientific CMOS (sCMOS) sensors have found increasing applications to X-ray detection, including X-ray astronomical observations. In order to examine the performance of sCMOS sensors, we have developed X-ray cameras based on sCMOS sensors. Two cameras, CNX22 and CNX 66, have been developed using sCMOS sensors with a photosensitive area of 2 cm × 2 cm and 6 cm × 6 cm, respectively. The designs of the cameras are presented in this paper. The CNX22 camera has a frame rate of 48 fps, whereas CNX66 has a frame rate of currently 20 fps, that can be boosted to 100 fps in the future. The operating temperature of the sCMOS sensor can reach to -20°C for CNX22 and -30°C for CNX66 with a peltier cooler device. In addition to the commonly used mode of saving original images, the cameras provide a mode of real-time extraction of X-ray events and storage their information, which significantly reduces the requirement for data storage and offline analysis work. For both cameras, the energy resolutions can reach less than 200 eV at 5.9 keV using single-pixel events. These cameras are suitable for X-ray spectroscopy applications in laboratories and calibration for the space X-ray telescopes.

## 1. INTRODUCTION

Due to their high readout frame, low power consumption, and high circuit integration, scientific CMOS sensors have increasingly been applied to X-ray astronomy and nuclear physics. There have been extensive studies of their performance[1], [2], [3]. We have been studying the performance of sCMOS sensors for astronomical applications as X-ray detectors, using the sCMOS sensors produced by Gpixel Inc., including Gsense400BSI[4] and Gsense1516BSI. The energy resolutions of 187 eV and 180.1 eV at 5.90 keV were obtained for Gsense400BSI and Gsense1516BSI, respectively[5], [6], [7]. We also found and corrected the crosstalk phenomenon in the Gsense400BSI series[8], which came from a design defect of the column structure. Several groups have studied the quantum efficiency of Gsense400BSI sensors with different coatings and the results showed generally good agreement with the theoretical curve[9], [10], [11]. Harada et al.[11] presented a new design of the sensor SP3 based on Gsense400BSI, which has a reduced thickness of the entrance window to 5 nm. This new design makes the sCMOS sensor reaches more than 90% quantum efficiency in the photon energy range of 80-1000 eV.

To test these sCMOS sensors, we have developed two types of sCMOS X-ray cameras based on two different sCMOS sensors. In addition to the traditional camera's readout method of saving the raw images, a unique X-ray event extraction algorithm is developed and integrated into the camera. The energy and position of X-ray events can be extracted in real-time, which significantly reduces the data storage requirement and can run the sCMOS sensor in its full frame rate. The camera can be easily applied to various test environments. The mechanics, electronics and its DAQ are introduced in section 2 and the software interface and the performance of the cameras are shown in section 3. Conclusions are presented in section 4.

* lingzhixing@nao.cas.cn

## 2. DESIGN OF THE CMOS CAMERA

### 2.1 The sCMOS cameras structures

We have developed two types of sCMOS cameras, named CNX22 and CNX66. Camera CNX22 uses a back-side illuminated scientific CMOS sensor, Gsense400BSI[4], with a photosensitive area of 2 cm × 2 cm. It has 2048 × 2048 pixels with a pixel size of 11 μm × 11 μm. The frame rate reaches 48 fps in standard (STD) mode. Camera CNX66 is based on a large back-side scientific CMOS sensor, Gsense1516BSI[7], with a photosensitive area of 6 cm × 6 cm. It has 4096 × 4096 pixels with a pixel size of 15 μm × 15 μm. The current frame rate reaches to 20 fps which can be boosted to 100 fps in the future. The specific properties of these two cameras are shown in Table 1.

Table 1. The specific properties of camera CNX22 and camera CNX66.

|  | Camera CNX22 | Camera CNX66 |
|---|---|---|
| sCMOS sensor | Gsense400BSI | Gsense1516BSI |
| Photosensitive area | 22.528 mm × 22.528 mm | 61.44 mm × 61.44 mm |
| Number of pixels | 2048 × 2048 | 4096 × 4096 |
| Pixel size | 11 μm × 11 μm | 15 μm × 15 μm |
| Frame rate | 48 fps @ STD | 20 fps |
| Shutter format | Rotting shutter | Rotting shutter |
| Fixed pattern noise (FPN) | 2.5 $e^-$ at high gain | <5.0 $e^-$ at high gain |
| Readout noise | <2.0 $e^-$ at high gain | <5.0 $e^-$ at high gain |
| Dark current | 345 $e^-$/p/s @ ~30°C * | <0.02 $e^-$/pixel/s @ -30°C <br> <10 $e^-$/pixel/s @ 20°C |
| Photo-response non-uniformity (PRNU) | <2 % * | <2 % * |
| Energy resolution (using single pixel events) | ~190 eV @ 5.9keV | ~180 eV @ 5.9keV |
| Cooling Temperature | -20°C | -30°C |
| Overall size | 16 × 16 × 21 cm$^3$ | 20 × 20 × 25 cm$^3$ |
| Weight | <5 kg | <5 kg |

* These parameters are from official datasheet.

The mechanics of these two types of sCMOS cameras are shown in Figure 1. Each camera has a sCMOS sensor chamber, which is sealed to reach negative pressure, and an electronics chamber, just below the sensor chamber. A vacuum pump can be connected to the sCMOS sensor chamber to provide a vacuum environment for the sCMOS sensor. The sCMOS sensor is mounted by a zero-insert force socket and can be easily removed and changed. The sensor can also be welded to the sensor board directly according to the application environment. A cooling finger is installed behind the sCMOS sensor which can cool the sCMOS sensor down to -20°C for camera CNX22 and -30°C for camera CNX66. The cooling speed is limited to 3°C per minute to protect the sCMOS sensor. The camera provides a Cf100 flange connector to be used in different environments. A blank flange with a Be window can be installed to test the X-ray performance and a glass window can be installed to test the optical performance. The camera can work at a vacuum down to 10$^{-5}$ Pa. The electronics and the cooling Fan are integrated inside the electronics chamber. The overall sizes of these cameras are 16 × 16 × 21 cm$^3$ and 20 × 20 × 25 cm$^3$ respectively, and the weight of each camera is less than 5 kg.

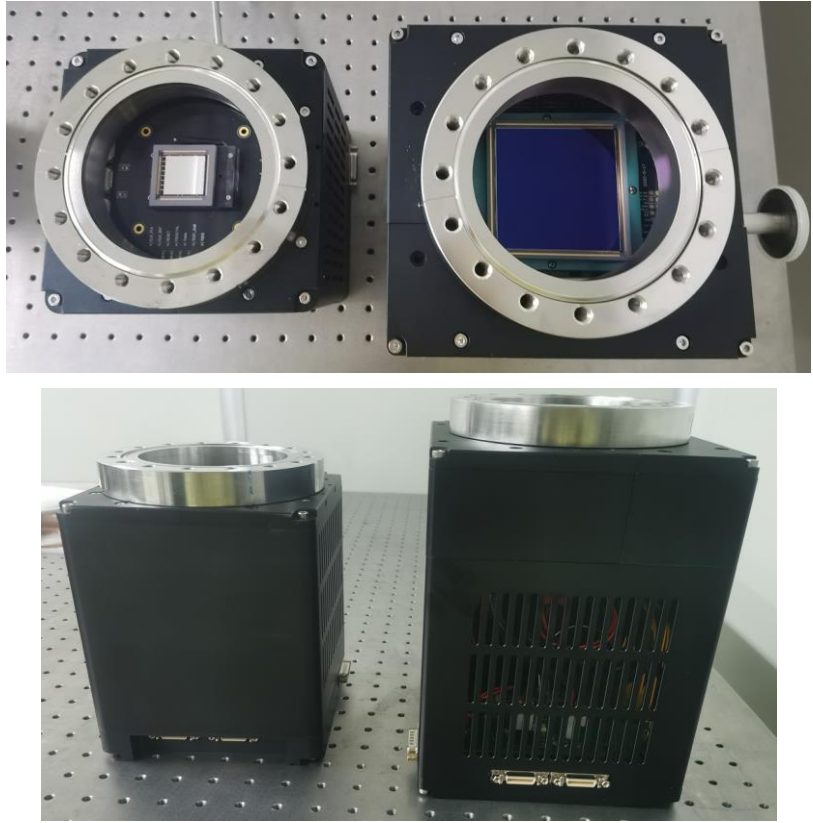

Figure 1. Camera CNX22 (left) and camera CNX66 (right).

## 2.2 The electronics and DAQ

The camera electronics include a sensor board, a peltier cooler device and an FPGA board. Figure 2 shows the schematic of the electronics. A 12V power is provided to the camera and then converted into the voltages required by the sCMOS sensor and the peltier cooler device respectively. The peltier control circuit provides the temperature control function to realize the driving and monitoring the temperature of the sCMOS sensor. The FPGA board provides clock and bias voltages for the sCMOS sensor and performs data processing and transmission. The data are transmitted to the computer through a standard cameralink bus.

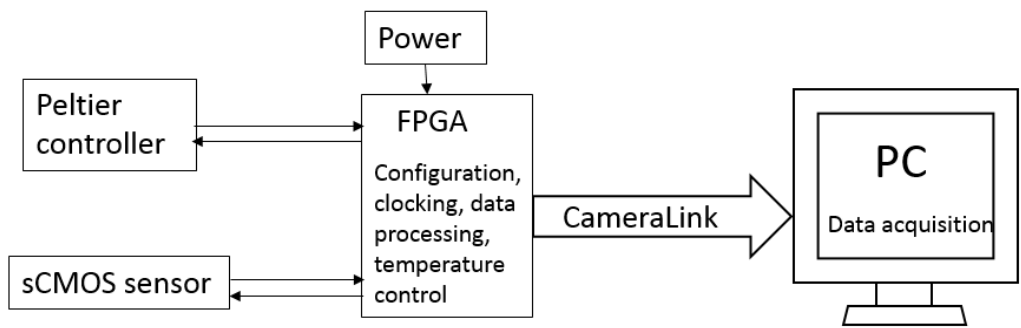

Figure 2. Schematic of the electronics.

## 2.3 Working modes of the camera

The camera provides two working modes: a raw image mode and an event-extraction mode. The raw image mode saves all the frame images from the camera. However, the saving speed is related to the hardware configuration. There will be some missing images when the camera runs at a high frame rate with a standard computer. For example, only 2 frame images per second can be saved for a standard computer when the camera CNX66 works at a frame rate of 20 Hz. The other 18 images would be discarded. The event-extraction method is a unique mode to record only the hitting area data of X-rays from the raw image. This can save the communication speed and the storage space effectively.

A bias map of the sCMOS sensor is required to extract X-ray events in the event-extraction mode. The bias map can be produced first according to the experiment setup. After subtracting the bias level, a search algorithm is operated to find regions of X-ray interaction location. Three sizes of regions for extracting an X-ray event are provided: 3 by 3 pixels, 5 by 5 pixels and 7 by 7 pixels. The choice of the region size depends on the performance of the sCMOS sensor itself. For a high-depleted sensor, a region of 3 by 3 pixels is suggested. For a low-depleted sensor, a 7 by 7 pixels region is needed to collect all of the electrons produced by an X-ray photon. The maximum local signal of the region is recorded as the hitting position of the X-ray event, and the sum of the whole region is recorded as the energy of the X-ray event. After recording the position, energy and time of X-ray events, the subsequent data analysis can be done in a straightforward way. Users also can only use the single pixel events, of which the electron produced by an X-ray photon is collected only by a single pixel. The energy resolution of the single pixel events is better than that of all events.

The event extraction mode can greatly reduce the requirements for data storage and one million X-ray events only need a storage size of 50MB. Currently, a maximum of 3000 X-ray events per image can be extracted in event-extraction mode. Taking camera CNX66 for example, an upper limit of 60000 counts/s can be recorded precisely. Compared to the traditional CCD X-ray sensors, this count rate makes a great improvement.

## 3. SOFTWARE AND PERFORMANCE

The software interface of the camera is shown in Figure 3. All of the configurations can be changed by a remote software directly and the camera can be controlled by a remote computer by a standard UDP protocol.

The operation mode of the raw image mode or event extraction mode can be easily chosen in the software. On the left panel, the configuration parameters, such as the integration time, temperature and gain, can be configured according to requirements. The software also provides functions to adjust the bias voltage and reset the power. The sCMOS temperature, the cooler temperatures of the hot and cold ends, as well as the power percentage of the cooler are monitored and displayed in the software directly.

When the X-ray event extraction mode is chosen, the bias map can be produced by the software or imported from a pre-generated file. The region of an X-ray can be set as 3 by 3 pixels, 5 by 5 pixels or 7 by 7 pixels. To take out the influence of the defect region of the sCMOS sensor, a mask image map can be loaded to remove the corresponding pixels or regions. For example, setting a mask image with the top 100 rows to true means that the X-ray extraction algorithm would not search the X-ray events in the top 100 rows. The minimum energy threshold should be set according to the test energy range. We suggest setting it to more than 10σ of the noise level. Otherwise, there will be a lot of false signals that will be mistaken as X-ray events.

The time, position and energy of each X-ray will be written into the storage disk during the experiment, as well as the operating temperature of the camera. In addition, some analysis results of X-rays are also displayed at the software interface and saved into the disk, such as the number of X-rays in each frame image, the cumulative image of all X-rays, their distribution in rows and columns and the spectrum of X-rays. Users can do the offline analysis easily.

The first sCMOS camera CNX22 was developed in 2018. Since then, a lot of experiments have been done with these cameras. Using single pixels, the energy resolution of less than 200 eV at 5.9 keV was obtained with camera CNX22 and CNX66 respectively. Figure 4 shows the spectrum of the $^{55}$Fe source tested with camera CNX66 at -30°C recently. The single-pixel events shown as the blue line account for about 45% of the total X-ray events. At 5.90 keV, the energy resolutions are 174 eV for the single-pixel events and 210 eV for all X-ray events. The energy resolutions obtained at different emission lines of elements are shown in Figure 5. For camera CNX66, the energy resolution at room temperature is even a little bit better than that at lower temperature[7]. Our recent study shows that after a correction algorithm developed, the energy resolution could reach 140 eV at 6400 eV at 20°C for single pixel events[12].

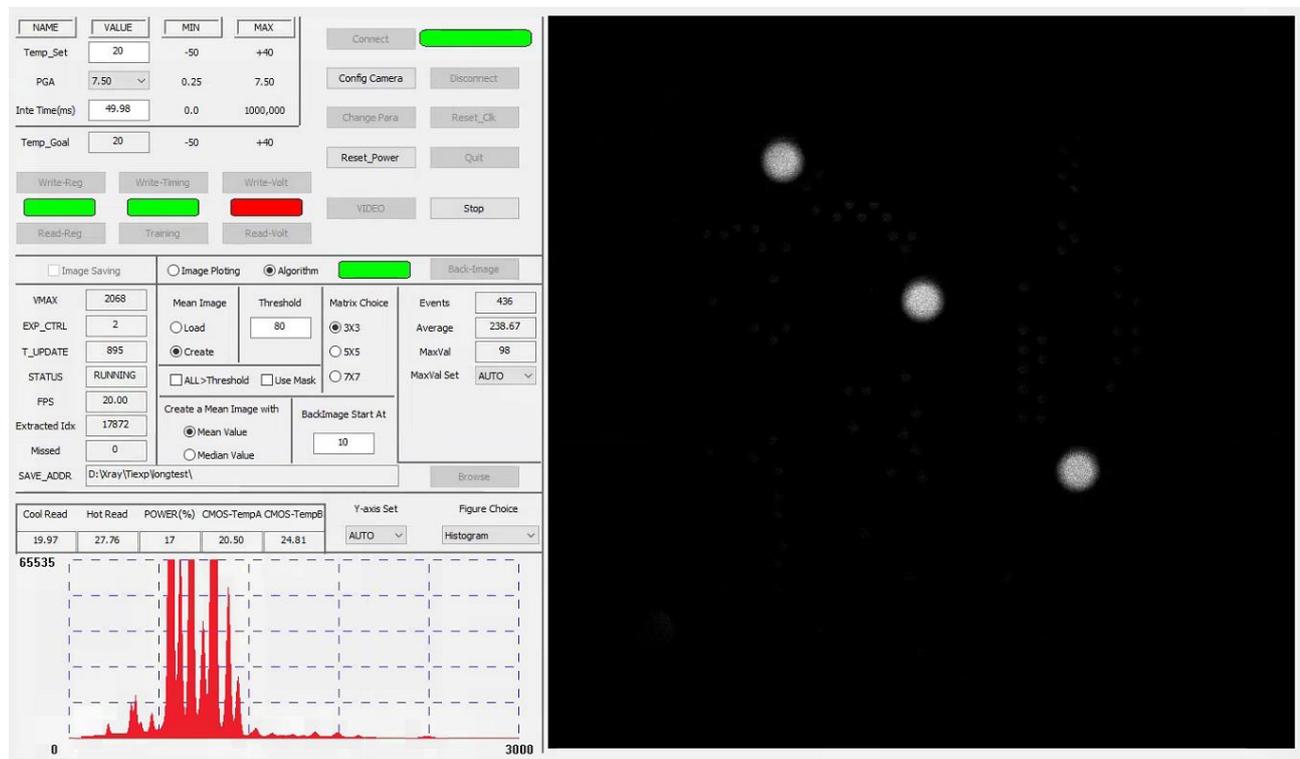

Figure 3. Camera software interface.

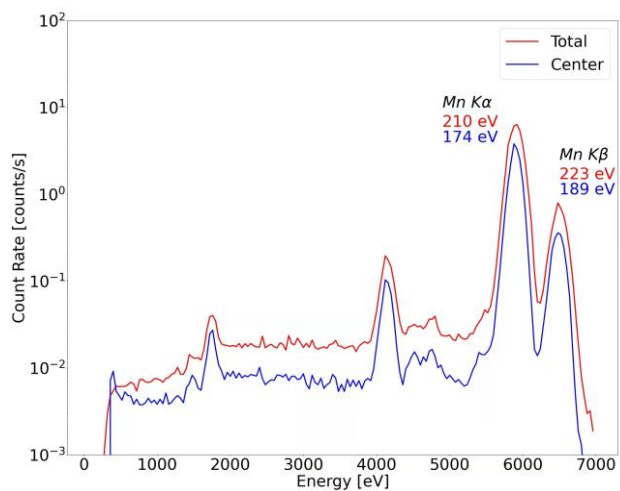

Figure 4. $^{55}$Fe spectrum tested with the camera CNX66.

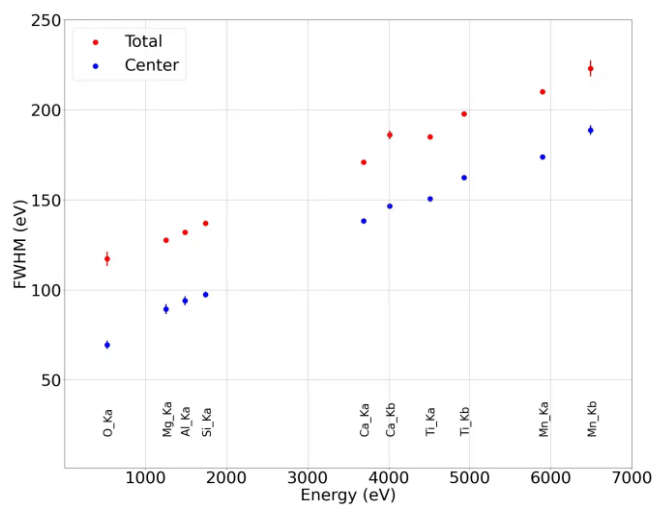

Figure 5. The energy resolutions obtained with different emission lines of elements taken by the camera CNX66.

## 4. CONCLUSION

We have developed two types of sCMOS cameras, CNX22 and CNX66, based on GSENSE400BSI and GSENSE1516BSI, respectively. The photosensitive areas are 2 cm × 2 cm and 6 cm × 6 cm respectively. The readout noises of the cameras are less than 2 electrons and 5 electrons. In order to satisfy different requirements of test environments, a CF100 flange connector is provided. Different windows can be mounted on the top of the camera to test the characteristic of the sCMOS sensor or test with visible light or X-rays. The camera can work in a vacuum environment of $10^{-5}$ Pa. The peltier cooler can cool the sCMOS sensor down to -20°C for camera CNX22 and -30°C for camera CNX66 in air condition. Two operation modes are provided: an image mode and an events extraction mode. Compared with storing the raw image, the X-rays extraction algorithm improves significantly the storage of valid data and reduces the requirement for storage. For both cameras, the energy resolution at 5.9 keV is around 200 eV. Our recent test result shows that the energy resolution of 174 eV at 5.90 keV was obtained with camera CNX66 using the single-pixel events at -30°C. The camera can be widely used in laboratories to detect visible light, UV and soft X-rays, as well as charged particles, such as electrons and protons. These make the sCMOS cameras very efficient and easy to use. In the future, we will further optimize and upgrade the functions and the software of the camera. A new vacuum camera is under development, which can be put inside the vacuum system as a whole.

## ACKNOWLEDGMENTS

This work is supported by the Strategic Pioneer Program on Space Science, Chinese Academy of Sciences, Grant No.XDA15310102-02 and the National Natural Science Foundation of China, Grant No. 12173055.

## REFERENCES

[1] Tanmoy Chattopadhyay, Abraham D. Falcone, David N. Burrows, et al., "X-ray Hybrid CMOS Detectors: Recent Development and Characterization Progress" Proc. of SPIE Vol. 10699, 106992E (2018)

[2] T Tanaka，TG Tsuru，H Uchida, et al., "Performance of SOI Pixel Sensors Developed for X-ray Astronomy", arXiv:1812.05803v1, 2018
[3] Noriyuki Narukage, Shin-nosuke Ishikawa, Taro Sakao, Xinyang Wang. "High-speed back-illuminated CMOS sensor for photon-counting-type imaging-spectroscopy in the soft X-ray range" Nuclear Inst. and Methods in Physics Research, A 950 (2020) 162974
[4] https://www.gpixel.com/products/area-scan-en/gsense/gsense400bsi-11-μm-4mp-rolling-shutter-image-sensor/.
[5] Wenxin Wang, Zhixing Ling, Chen Zhang, et al., "Developments of scientific CMOS as focal plane detector for Einstein probe mission", Proc. SPIE 10699 (2018) 106995O.
[6] W.X. Wang, Z.X. Ling, C. Zhang, et al. "Characterization of a BSI sCMOS for soft X-ray imaging spectroscopy" 2019 JINST 14 P02025
[7] Qinyu Wu, Zhenqing Jia, Wenxin Wang, et al. "X-Ray Performance of a Customized Large-format Scientific CMOS Detector" Publications of the Astronomical Society of the Pacific, 134:035006 (15pp), 2022 March
[8] Z. Ling, W.Wang, Z. Jia, et al., "A correlogram method to examine the crosstalk of sCMOS sensors" JInst, 2021,16, P03018
[9] Kewin Desjardins, Horia Popescu, Pascal Mercère, et al., "Characterization of a back-illuminated CMOS Camera for soft x-ray coherent scattering" AIP Conference Proceedings 2054, 060066 (2019)
[10] Tetsuo Harada, Nobukazu Teranishi, Takeo Watanabe, et al., "Energy- and spatial-resolved detection using a backside-illuminated CMOS sensor in the soft X-ray region" Applied Physics Express 12, 082012 (2019)
[11] Tetsuo Harada, Nobukazu Teranishi, Takeo Watanabe, et al., "High-exposure-durability, high-quantum-efficiency (>90%) backside-illuminated soft-X-ray CMOS sensor" Applied Physics Express 13, 016502 (2020)
[12] Ling et al. in preparation.